\documentclass[a4paper,useAMS,usenatbib]{mn2e}
\usepackage[OT1]{fontenc}
\usepackage[latin9]{inputenc}
\usepackage{amsmath}
\usepackage{graphicx}
\usepackage[authoryear]{natbib}

\makeatletter

\newcommand{\noun}[1]{\textsc{#1}}
\providecommand{\tabularnewline}{\\}

\usepackage{esint}

\newcommand{\hompc}{\,h\,{\rm Mpc}^{-1}}
\newcommand{\mpcoh}{\,h^{-1}\,{\rm Mpc}}
\newcommand{\simgt}{\,\hbox{\lower0.6ex\hbox{$\sim$}\llap{\raise0.6ex\hbox{$>$}}}\,}
\newcommand{\simlt}{\,\hbox{\lower0.6ex\hbox{$\sim$}\llap{\raise0.6ex\hbox{$<$}}}\,}
\usepackage{aas_macros}

\usepackage{bm}

\title[Neutrino masses from red and blue galaxies]{Neutrino masses from clustering of red and blue galaxies: a test of astrophysical uncertainties}
\date{\today}
\author[M. E. C. Swanson et al.]{Molly E. C. Swanson,$^{1,2}$\thanks{E-mail: molly@star.ucl.ac.uk}, Will J. Percival,$^{3}$ and Ofer Lahav$^{1}$ \\  $^{1}$Department of Physics and Astronomy, University College London, Gower Street, London, WC1E 6BT, UK\\
$^{2}$Harvard-Smithsonian Center for Astrophysics, 60 Garden Street, Cambridge, MA 02138, USA\\   
$^{3}$Institute of Cosmology and Gravitation, University of Portsmouth, Dennis Sciama building, Portsmouth, P01 3FX, UK} 

\makeatother

\begin{document}
\maketitle

\begin{abstract}
Combining measurements of the galaxy power spectrum and the cosmic
microwave background (CMB) is a powerful means of constraining the
summed mass of neutrino species $\sum m_{\nu}$, but is subject to
systematic uncertainties due to non-linear structure formation, redshift-space
distortions and galaxy bias. We empirically test the robustness of
neutrino mass results to these effects by separately analyzing power
spectra of red and blue galaxies from the Sloan Digital Sky Survey
(SDSS-II) Data Release 7 (DR7), combined with the CMB five-year Wilkinson
Microwave Anisotropy Probe (WMAP5) data. We consider fitting for a
range of maximum wavenumber $k$ using twelve different galaxy bias
models. For example, using a new model based on perturbation theory
and including redshift space distortions \citep{2009PhRvD..80h3528S},
the all-galaxy power spectrum combined with WMAP5 for a wavenumber
range of $k<0.2\hompc$ yields 95\% CL $\sum m_{\nu}<0.46$ eV. The
red and blue galaxy power spectra give 0.41 and 0.63 eV respectively
for this model. Using mock catalogues, we find the expected difference
in these limits assuming a true neutrino mass of zero is $0.10\pm0.14$
eV. Thus the difference of 0.22 eV between upper limits on neutrino
mass for red and blue galaxies is approximately $1\sigma$ from the
expected value. We find similar results for the other models and $k$
ranges tested. This indicates good agreement for current data but
hints at possible issues for next-generation surveys. Being able to
perform such systematic tests is advantageous, and future surveys
would benefit by including broad galaxy populations and luminosities
that enable such a decomposition. 
\end{abstract}

\section{Introduction}

Atmospheric and solar neutrino experiments show that neutrinos have
mass and that there is significant mixing between the various neutrino
interaction eigenstates \citep{1996NuPhS..48..284D,1998PhRvL..81.1562F.good,2002PhRvL..89a1301A.good,2005PhRvD..71k2005A.good,2009arXiv0905.4907M,2010PhRvD..81i2004W.good}.
Cosmological measurements nicely complement these experiments by constraining
the sum of the masses of the different eigenstates $\sum_{i}m_{\nu_{i}}$.
The cosmological experiments exploit differences between the behaviour
of cold dark matter (CDM) and massive neutrinos, which free-stream
when relativistic, removing small-scale structure that would have
formed in a CDM-only universe (for recent reviews see, e.g., \citealt{2005NJPh....7...61E,2006PhR...429..307L,2010arXiv1007.0658H}).
To first order, the suppression depends on $\Omega_{\nu}$, the total
mass density comprised of neutrinos relative to the critical density
of the Universe. This is related to the sum of the mass eigenstates
by $\sum m_{\nu}=\Omega_{\nu}\left(94.1\, h^{2}\mathrm{eV}\right)$,
where $h$ is the Hubble constant $H_{0}$ divided by $100\,\mathrm{km\, s^{-1}Mpc^{-1}}$.
Thus cosmological measurements primarily probe the total particle
mass summed over neutrino species. In principle, the neutrino mass
hierarchy could be measured by resolving the free-streaming scale
of individual neutrino species, but current experiments do not have
the sensitivity required to do this \citep{2003PhRvD..67h5017H,2006PhRvD..73h3520T,2010JCAP...05..035J}.

The most recent upper limit on the summed neutrino mass published
by the Wilkinson Microwave Anisotropy Probe (WMAP) team is $\sum m_{\nu}<0.58\,\mathrm{eV}$
at 95\% CL \citep{2010arXiv1001.4538K.good}.This limit combines CMB
data from the 7 year WMAP data release (WMAP7) with a measurement
of $H_{0}$ from type Ia supernovae (SN Ia) and the Baryon Acoustic
Oscillation (BAO) information from the 2-degree Field Galaxy Redshift
Survey (2dFGRS) and Sloan Digital Sky Survey (SDSS-II). Adding large
scale structure information from surveys of galaxies, galaxy clusters,
or weak lensing pushes this limit down even further: other groups
have combined WMAP5 with SN Ia, BAO, and large scale structure information
and derive limits as low as $\sum m_{\nu}<0.28\,\mathrm{eV}$ at 95\%
CL (e.g., \citealt{2009ApJ...692.1060V,2009A&A...500..657T,2010JCAP...01..003R,2010MNRAS.406.1805M,2010PhRvL.105c1301T}).
The challenge now is to reliably bring down the upper limits to the
$0.1\,\mathrm{eV}$ level or even detect the neutrino mass using the
next generation of galaxy survey and CMB data \citep{2004PhRvD..70d5016L,2006PhR...429..307L,2007JCAP...07..004H,2010MNRAS.405..168L}.

Unfortunately, our ability to measure the summed neutrino mass using
large scale structure information such as the galaxy power spectrum
will likely be limited by our lack of knowledge about other effects
that change the expected amplitude of galaxy clustering on small-scales.
These effects include the formation of non-linear matter structures
and the non-linear galaxy peculiar velocities within these structures.
The peculiar velocities affect the measured power spectrum because
we infer distances from redshifts, and peculiar velocities are misinterpreted
as being due to the Hubble flow. Thus the random motion of galaxies
within collapsed structures acts to damp the power spectrum on small
scales.

In addition, galaxies are not expected to exactly Poisson sample the
distribution of matter in the Universe \citep{1999MNRAS.304..767S,2002MNRAS.333..730C};
evidence for this has been seen in, e.g., \citet{2008MNRAS.387.1045W}.
This severely complicates the extraction of cosmological information
from galaxy surveys. Furthermore, it has been known for some time
that different populations of galaxies demonstrate different clustering
strengths \citep{1976ApJ...208...13D,1980ApJ...236..351D,1994ApJ...431..569P,1994MNRAS.267.1020P,1999MNRAS.309...89S,2001MNRAS.328...64N.good,2002MNRAS.332..827N.good,2002ApJ...571..172Z.good,2005ApJ...621...22Z,2005MNRAS.356..247W.good,2005MNRAS.361..415C,2006MNRAS.368...21L,2008MNRAS.385.1635S},
showing that they cannot all have a simple relationship linking their
distribution with that of the matter. This galaxy bias (the relationship
between the galaxies and the matter) is the most pernicious physical
effect that limits our ability to extract cosmological data from galaxy
surveys \citep{2007ApJ...657..645P,2008MNRAS.385..830S}.

Because of the uncertainty regarding these effects, many cosmologists
(including the WMAP team) choose not to use the galaxy power spectrum
in making their cosmological parameter constraints \citep{2009ApJS..180..306D,2010arXiv1001.4538K.good}
and instead encapsulate the information from galaxy surveys as a prior
on the geometrical distance to the mean redshift of the galaxy survey
based on using the BAO acoustic peak as a standard ruler. This is
effective but discards a great deal of information contained in the
shape of the power spectrum that is especially valuable for making
neutrino mass constraints. In order to fully exploit the information
contained in the galaxy distribution, we must develop a more sophisticated
understanding of how the power spectrum is affected by galaxy bias,
non-linear evolution, and redshift space distortions. Of these three
systematic effects, the galaxy bias is in some sense the most troublesome:
non-linear evolution and redshift space distortions can be studied
in N-body simulations allowing only for evolution through gravity,
but modelling galaxy bias in detail requires hydrodynamic simulations
or semi-analytic galaxy formation models.

The aim of this paper is to quantify the impact of these systematic
effects on measurements of the summed neutrino mass from the galaxy
power spectrum, focusing especially on the effects of galaxy bias.
We use data from the final data release of the SDSS-II survey, which
we split into the bimodal populations of blue and red galaxies. By
comparing results from two galaxy populations with differing properties,
we can study the impact of astrophysical effects while bypassing the
need to model biasing in detail. We also consider how the range of
wavenumber $k$ fitted affects the results. Using data at larger $k$
values adds more information about neutrinos from the effects on the
small-scale structure, but it comes at the price of requiring more
complicated theoretical modelling. We fit our subsamples and the combined
sample with 12 different power spectrum models and combine with CMB
constraints from WMAP5.

This work is timely for two reasons. Firstly, the upcoming generation
of galaxy surveys such as the Dark Energy Survey (DES; \citealt{2006SPIE.6267E.119A.good}),
the Panoramic Survey Telescope \& Rapid Response System (Pan-STARRS;
\citealt{2008SPIE.7012E..54M.good,2008SPIE.7012E..95M.good}), and
the Large Synoptic Survey Telescope (LSST; \citealt{2009arXiv0912.0201L.good})
will significantly reduce the statistical uncertainty on power spectrum
measurements, making a detailed understanding of the systematic uncertainties
increasingly essential. Secondly, we can now make use of new perturbation
theory results that provide more accurate models for the non-linear
matter power spectrum including neutrinos \citep{2009PhRvD..80h3528S,2009PhRvD..80l3503T}:
along with \citet{2010arXiv1006.4845S} this analysis is one of the
first applications of these models to observational data.

This paper is organised as follows. We start by describing the data
that we use in Section~\ref{sec:data}. The method by which these
data are analysed, including power spectrum calculation and likelihood
analysis is described in Section~\ref{sec:method}. Results are presented
in Section~\ref{sec:results}, and are discussed in Section~\ref{sec:conclusions}.
For simplicity, here we assume a cosmological model of flat $\Lambda$CDM
plus massive neutrinos, but we note that alternative cosmological
models (in particular those with dark energy equation of state $w\neq-1$)
can impact neutrino mass measurements as well \citep{2005PhRvL..95v1301H,2008PhRvD..77f3005K,2010JCAP...01..003R}
and will be investigated further in future work. Specifically, we
fix $\Omega_{M}+\Omega_{\Lambda}=1$, the dark energy equation of
state $w=-1$, the tensor amplitude and running of the spectral index
to zero, and the effective number of relativistic degrees of freedom
$N_{\mathrm{eff}}=3.04$. We convert galaxy redshifts to distances
using $\Omega_{M}=0.25$ and $\Omega_{\Lambda}=0.75$, and assume
that this is the true cosmology when estimating errors on these measurements.

\section{Data}

\label{sec:data}

\subsection{Galaxy data }

The SDSS-II experiment used a 2.5m telescope \citep{2006AJ....131.2332G.good},
to obtain a spectroscopic sample of galaxies selected to a limiting
Galactic extinction-corrected Petrosian magnitude $r<17.77$, or $r<17.5$
in a small subset of the early data from the survey \citep{2002AJ....124.1810S.good}.
There are approximately $90$ galaxies per square degree, with a median
redshift of $z=0.1$. The DR7 sample \citep{2009ApJS..182..543A.good}
used in our analysis includes $669\,905$ main galaxies \citep{2002AJ....124.1810S.good}.
We correct for fibre packing problems where close galaxies cannot
both be observed by assigning the redshift of the nearest observed
galaxy to a galaxy which was not observed, matching the methodology
of \citet{2002ApJ...571..172Z.good}. This is the subsample of SDSS
main galaxies used by \citet{2010MNRAS.401.2148P.good}. Galaxy redshifts
were converted to distances using our fiducial cosmology (flat $\Lambda$CDM
model with $\Omega_{m}=0.25$).

We have $k$-corrected galaxy luminosities using the methodology outlined
in \citet{2003AJ....125.2348B,2003AJ....125.2276B}. We also use a
$z=0.1$ shifted $r$-band filter to define our luminosities (as discussed
in \citealt{2003AJ....125.2348B,2003AJ....125.2276B}), which we refer
to as M$_{^{0.1}r}$. Absolute magnitudes and $k$-corrections were
calculated assuming $H_{0}=100\,{\rm km\, s}^{-1}{\rm Mpc}^{-1}$,
and our fiducial cosmology, and we have applied the recommended AB
corrections to the observed SDSS magnitude system \citep{2002AJ....123.2121S.good}.

This sample of galaxies was split into red and blue subsamples defined
by a constant colour cut of $M_{0.1_{g}}-M_{0.1_{r}}=0.8$. This approximately
divides the bimodal distribution of galaxies into red and blue types
(e.g. \citealt{2009MNRAS.392..682C}). Our results are not sensitive
to the exact nature of this division. The red and blue subsamples
have median redshifts of $z=0.111$ and $z=0.085$ respectively.

\subsection{CMB data}

\label{sub:cmb}

\begin{figure}
\includegraphics[width=1\columnwidth]{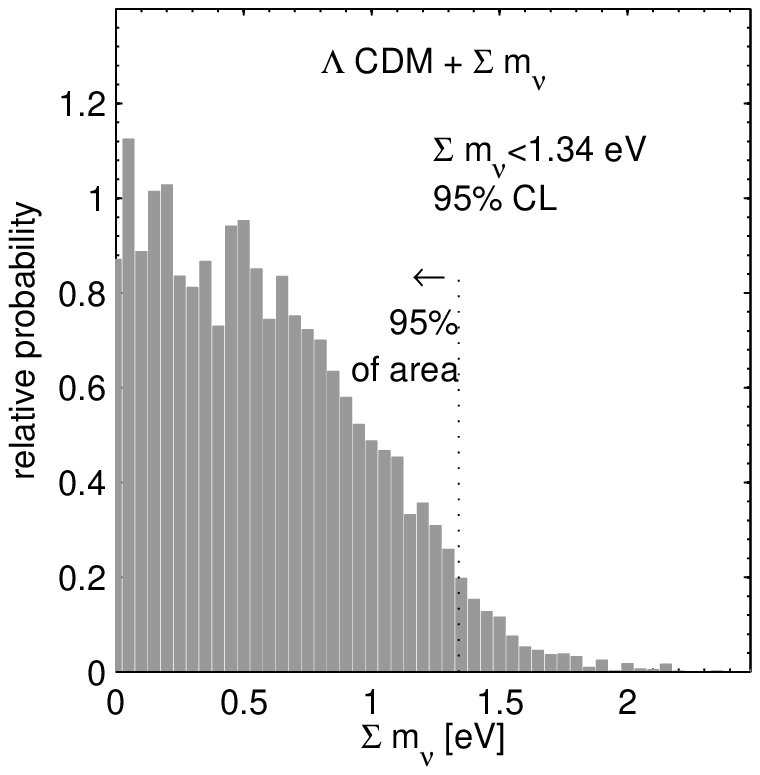}

\caption{\label{fig:mnuhist_wmap5}Histogram of the probability distribution
for $\sum m_{\nu}$ reproduced using the WMAP5 MCMC chain from \protect\citet{2009ApJS..180..306D}.}

\end{figure}

We use the WMAP5 CMB data as our baseline for neutrino mass constraints
\citep{2009ApJS..180..306D}. The WMAP5 CMB data alone provide a limit
of $\sum m_{\nu}<1.3\,\mathrm{eV}$ (95\% CL), robust to within 10\%
to varying tensor amplitude, running spectral index, or dark energy
equation of state $w$. As discussed in \citet{2005PhRvD..71d3001I},
the best one can hope to do for the neutrino mass limit using only
CMB data is $\sum m_{\nu}\la1.2\,\mathrm{eV}$. Since the WMAP5 limit
is already nearly at this value, we do not expect significant improvement
by going to WMAP7. This is borne out by the fact that the WMAP7-only
limit is also $\sum m_{\nu}<1.3\,\mathrm{eV}$\citep{2010arXiv1001.4538K.good}.

We have downloaded the WMAP5 $\Lambda$CDM+$M_{\nu}$ MCMC chain from
\citet{2009ApJS..180..306D} from the LAMBDA archive%
\footnote{\texttt{http://lambda.gsfc.nasa.gov}%
}. This model uses eight parameters to fit to the CMB temperature and
polarization power spectra: the cold dark matter density $\Omega_{c}h^{2}$,
the baryon density $\Omega_{b}h^{2}$, the neutrino mass density $\Omega_{\nu}h^{2}$,
the dark energy density $\Omega_{\Lambda}$, the amplitude of curvature
fluctuations $\Delta_{\mathcal{R}}^{2}$, the scalar spectral index
$n_{s}$, the reionization optical depth $\tau$, and the amplitude
$A_{\mathrm{SZ}}$ of a Sunyaev-Zel'dovich (SZ, \citealt{1970Ap&SS...7....3S})
fluctuation template spectum. We express these parameters in a more
digestible form using the flatness assumption, $\sum m_{\nu}=\Omega_{\nu}\left(94.1\, h^{2}\mathrm{eV}\right)$,
and the definition of the linear theory density fluctuation amplitude
at $8\mpcoh$ scales: \begin{equation}
\sigma_{8}\equiv\left\{ \frac{1}{2\mathrm{\pi}^{2}}\int_{0}^{\infty}\left[3\left(\sin x-x\cos x\right)/x^{3}\right]^{2}P_{\mathrm{lin}}\left(k\right)k^{2}dk\right\} ^{1/2}\label{eq:sig8}\end{equation}
 with $x\equiv k\times8\mpcoh$. This gives a parameter set $\bm{\theta}^{i}\equiv\left(\Omega_{\Lambda},\,\Omega_{b}/\Omega_{m},\, h,\,\sigma_{8},\, n_{s},\,\sum m_{\nu}\right)$
plus the CMB-only parameters $\left(\tau,\, A_{\mathrm{SZ}}\right)$
with $\Omega_{m}\equiv\Omega_{c}+\Omega_{b}+\Omega_{\nu}$.

Using this chain we have reproduced the $\sum m_{\nu}<1.3\,\mathrm{eV}$
limit, illustrated in Fig.~\ref{fig:mnuhist_wmap5}. This is a histogram
of the $N=7503$ values of $\sum m_{\nu}\equiv M_{\nu}$ in the MCMC
chain weighted by the weights $w_{\mathrm{CMB}}$ given in the chain.
The 95\% confidence limit on $\sum m_{\nu}$, which we denote as $M_{\nu}^{\mathrm{lim}}$
, is defined in the usual way as a region enclosing 95\% of the histogram
area: \begin{equation}
.95=\frac{\sum_{M_{\nu}^{i}<M_{\nu}^{\mathrm{lim}}}w_{\mathrm{CMB}}^{i}}{\sum_{i=1}^{N}w_{\mathrm{CMB}}^{i}},\label{eq:wmap_only}\end{equation}
 where $i$ is an index running through the entries in the MCMC chain.

\section{Analysis method}

\label{sec:method}

We explore the robustness of a neutrino mass limit produced by combining
the CMB with information from the galaxy power spectrum. We do this
by performing the analysis using power spectra of different subsamples
of galaxies, namely red and blue galaxies, and determining whether
the resulting limits are consistent with each other, in the spirit
of \citet{2008MNRAS.385..830S}. We also test the effect of changing
the range of $k$ values used from the galaxy power spectra and using
different prescriptions for the theoretical modeling of the nonlinear
power spectra.

\subsection{Calculating power spectra and uncertainties}

\label{sub:powerspectraandmocks}

Power spectra were calculated for the red, blue and combined catalogs
using the Fourier method of \citet{1994ApJ...426...23F}, as applied
by \citet{2010MNRAS.401.2148P.good}. In this method a weighted galaxy
over-density field is defined and Fourier transformed, then the spherically
averaged power is measured. The shot noise due to the discreteness
of the galaxy locations is subtracted, assuming the galaxies are a
Poisson sampling of the density distribution. We use the luminosity
(but not colour) dependent galaxy weights advocated by \citet{2004MNRAS.347..645P}.
These will at least partially correct for power spectrum shape changes
caused by variations in the galaxy bias over the redshift range of
the survey due to flux-limited selection and galaxy evolution.

The power spectrum for the combined catalog is not independent of
the red and blue catalogs since it includes the same galaxies. The
all-galaxy power spectrum $P_{\mathrm{all}}\left(k\right)$ is related
to the red and blue galaxy power spectra $P_{\mathrm{rr}}\left(k\right)$
and $P_{\mathrm{bb}}\left(k\right)$ by\[
P_{\mathrm{all}}\left(k\right)=f_{r}^{2}P_{\mathrm{rr}}\left(k\right)+2f_{r}f_{b}P_{\mathrm{rb}}\left(k\right)+f_{b}^{2}P_{\mathrm{bb}}\left(k\right)\]
 where $f_{r}$ and $f_{b}$ are the fractions of red and blue galaxies
in the sample ($f_{r}+f_{b}=1$) and $P_{rb}\left(k\right)$ is the
cross power spectrum between red and blue galaxies. Thus $P_{\mathrm{all}}\left(k\right)$
contains the information from the two subsample power spectra plus
the cross power spectrum.

Because of the survey mask (both angular and radial), the measured
power spectrum is a convolution of the true power spectrum \citep{1973ApJ...185..757H,1980lssu.book.....P.good,2001MNRAS.327.1297P.good}.
A discretized window function was used to quickly perform this convolution,
as described in \citet{2007ApJ...657..645P}. In order to calculate
the covariances between the power spectrum band-powers, we have created
10\,000 Log-Normal (LN) density fields \citep{1999ApJ...521L...5C,2005MNRAS.362..505C.good}
with a power spectrum for a flat $\Lambda$CDM cosmology with $\bm{\theta}^{i}=$($\Omega_{\Lambda}=0.7$,
$\Omega_{b}/\Omega_{m}=0.15$, $h=0.7$, $n_{s}=1$, $\sum m_{\nu}=0$),
and normalisation matched to the amplitude of the measured power.

From each LN density field $\delta\left(\bm{x}\right)$, we draw red,
blue, and combined galaxy catalogues as Poisson samplings matched
to the galaxy bias and spatial distribution of the SDSS-II samples.
That is, the mean $\lambda\left(\bm{x}\right)$ of the Poisson distribution
is given by $\lambda\left(\bm{x}\right)=\bar{n}\left(\bm{x}\right)\left(1+b\delta\left(\bm{x}\right)\right)$,
where $\bar{n}\left(\bm{x}\right)$ is the mean density including
the effects of the survey mask and $b$ is the linear galaxy bias
at large scales ($k<0.1\hompc$) for the red and blue galaxy samples.
Catalogues were calculated on a $(512)^{3}$ grid with box length
$4000\mpcoh$. Unlike N-body simulations, these mock catalogues do
not model the growth of structure, but instead return a density field
with a log-normal distribution, similar to that seen in the real data.
The mock catalogues also do not include redshift space distortions.
Power spectra for the mock catalogues were calculated following the
same method used for the SDSS-II data and were used to compute covariance
matrices for each subsample.

A subset of 200 of the mock catalogue power spectra was also used
to estimate the expected sensitivity of the neutrino mass limit. We
do this by applying our analysis described in Section~\ref{sub:Importance-sampling}
to the mock power spectra and define a range using the mean $\pm$
one standard deviation of the 95\% CL limits on $\sum m_{\nu}$ from
the mocks. Because the mock catalogues have $\sum m_{\nu}=0$ and
linear bias, the expected sensitivity range is the range in which
we would expect to measure the neutrino mass limit assuming that the
neutrino mass were equal to zero and that we understood bias perfectly.
This provides a baseline for interpreting the neutrino mass limits
determined from the actual data.

\subsection{Models for galaxy bias and nonlinear evolution}

\label{sub:bias}

\begin{figure*}
\includegraphics[width=1\textwidth]{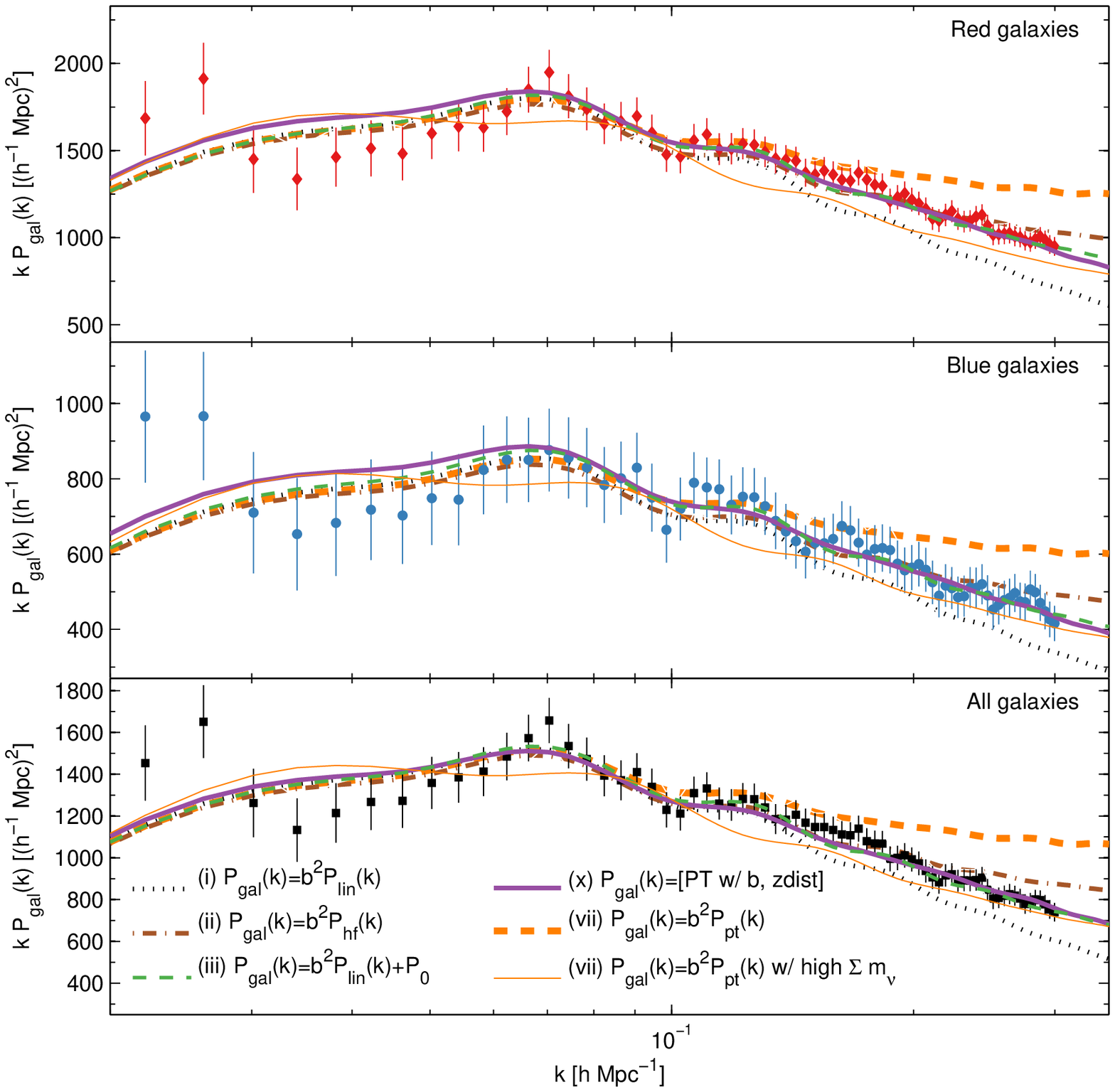}

\caption{\label{fig:pktests}Measured SDSS-II DR7 power spectra for red (top
panel), blue (middle panel), and all (bottom panel) galaxies compared
to several theoretical models. All curves plotted with thick lines
use the cosmology from the WMAP5 $\Lambda$CDM+$\sum m_{\nu}$ MCMC
chain that yields the maximum likelihood for WMAP5 only. Models \ref{P}
and \ref{linz} are plotted with the bias parameters that give the
best fit to the $P\left(k\right)$ data shown for $k<0.3\, h\mathrm{Mpc}^{-1}$
and models \ref{lin}, \ref{smith}, and \ref{saito} are plotted
with the best-fit $b$ for the data shown at $k<0.1\, h\mathrm{Mpc}^{-1}$
to better illustrate the different behaviours at small scales. The
thin orange line and the thick dashed orange line both use model \ref{saito},
but the thin orange line uses cosmology and bias parameters that give
the maximum likelihood to the $P\left(k\right)$ data at $k<0.3\, h\mathrm{Mpc}^{-1}$
for the $P_{\mathrm{gal}}\left(k\right)=b^{2}P_{\mathrm{pt}}\left(k\right)$
model -- this cosmology has a high value of with a high $\sum m_{\nu}$
and fits the data better than the thick dotted orange line (though
not as well as models \ref{P} and \ref{linz}). This illustrates
how increasing the neutrino mass can lead to a better fit for some
models. See text for more detailed discussion.}

\end{figure*}

Our starting point for theoretical modeling is the linear power spectrum
$P_{\mathrm{lin}}\left(k,z\right)$ calculated using \noun{camb}%
\footnote{\texttt{http://camb/info/}%
} for our set of cosmological parameters $\bm{\theta}^{i}\equiv\left(\Omega_{\Lambda},\,\Omega_{b}/\Omega_{m},\, h,\,\sigma_{8},\, n_{s},\,\sum m_{\nu}\right)$.
We evaluate $P_{\mathrm{lin}}\left(k,z\right)$ at the median redshift
of each galaxy sample ($z=0.111$ for red, $z=0.085$ for blue, and
$z=0.099$ for all galaxies) and drop the explicit dependence on $z$
in our power spectra for notational convenience.

We consider twelve different models for converting $P_{\mathrm{lin}}\left(k,z\right)$
into a galaxy power spectrum. The first six models represent the typical
models used in previous cosmological analyses of the neutrino mass
and have either one or two additional free parameters, $b$ and a
free parameter that controls the amplitude of the non-linear change
in shape between galaxy and linear matter power spectrum, which can
take on different values for different galaxy subsamples.

For our remaining six models, we explore variations on new models
for the matter power spectrum including a contribution from massive
neutrinos, based on perturbation theory described in \citet{2008PhRvL.100s1301S,2009PhRvD..80h3528S,2009PhRvD..80l3503T}.
For this theory, the baseline model is \begin{equation}
P_{\mathrm{pt}}\left(k\right)\equiv f_{\mathrm{cb}}P_{\mathrm{1loop}}^{\mathrm{cb}}\left(k\right)+2f_{\mathrm{cb}}f_{\nu}P_{\mathrm{lin}}^{\mathrm{cb}\nu}\left(k\right)+f_{\nu}^{2}P_{\mathrm{lin}}^{\nu}\left(k\right),\label{eq:saito}\end{equation}
 where $f_{\mathrm{cb}}$ and $f_{\nu}$ are the mass fractions of
cold dark matter plus baryons and neutrinos respectively, $P_{\mathrm{lin}}^{\nu}\left(k\right)$
is the linear theory power spectrum for neutrinos only, $P_{\mathrm{lin}}^{\mathrm{cb}\nu}\left(k\right)$
is the linear theory cross power spectrum between cold dark matter
plus baryons and neutrinos, and $P_{\mathrm{1loop}}^{\mathrm{cb}}\left(k\right)$
is the nonlinear power spectrum of cold dark matter plus baryons computed
to one loop in perturbation theory, which has been studied extensively
in the literature (e.g., \citealt{1981MNRAS.197..931J,1992PhRvD..46..585M,1994ApJ...431..495J,1996ApJS..105...37S,1998MNRAS.301..797H}).
Here we use \begin{equation}
P_{\mathrm{1loop}}^{\mathrm{cb}}\left(k\right)\equiv P_{\mathrm{lin}}^{\mathrm{cb}}\left(k\right)+P_{\mathrm{\mathrm{(22)}}}^{\mathrm{cb}}\left(k\right)+P_{\mathrm{(13)}}^{\mathrm{cb}}\left(k\right),\label{eq:1loop}\end{equation}
 where $P_{\mathrm{\mathrm{(22)}}}^{\mathrm{cb}}\left(k\right)$ and
$P_{\mathrm{(13)}}^{\mathrm{cb}}\left(k\right)$ are the approximations
to the one-loop correction spectra given by equations~(32)~and~(36)
in \citet{2009PhRvD..80h3528S}. If we replace $P_{\mathrm{1loop}}^{\mathrm{cb}}\left(k\right)$
by $P_{\mathrm{lin}}^{\mathrm{cb}}\left(k\right)$ in equation~\eqref{eq:saito},
the expression for $P_{\mathrm{pt}}\left(k\right)$ reduces to $P_{\mathrm{lin}}\left(k\right)$.
The motivation for this model is that while cold dark matter and baryon
perturbations start to behave non-linearly at $k\ga0.1\, h\mathrm{Mpc}^{-1}$
, perturbations of the neutrino density will tend to remain mostly
linear due to the large velocity dispersion of the neutrinos. Similar
models have been explored by \citet{2006JCAP...06..025H,2009PhRvD..79b3520I}
and further improved by \citet{2008JCAP...10..035W,2009JCAP...06..017L}.

The twelve models used in this analysis are as follows:

\begin{enumerate}
\item The simplest model for the galaxy power spectrum is that of constant
linear bias of this linear power spectrum: \begin{equation}
P_{\mathrm{gal}}\left(k\right)=b^{2}P_{\mathrm{lin}}\left(k\right),\label{eq:lin}\end{equation}
 where the bias $b$ is allowed to take on different values for different
galaxy subsamples. \label{lin} 
\item Our second model replaces the linear matter power spectrum with the
Halofit nonlinear transformation \citep{2003MNRAS.341.1311S}, which
we denote $P_{\mathrm{hf}}\left(k\right)$: \begin{equation}
P_{\mathrm{gal}}\left(k\right)=b^{2}P_{\mathrm{hf}}\left(k\right).\label{eq:smith}\end{equation}
 \label{smith} 
\item For the next model we add one more free parameter to model~\ref{lin}
as an additive constant term: \begin{equation}
P_{\mathrm{gal}}\left(k\right)=b^{2}P_{\mathrm{lin}}\left(k\right)+P_{0}.\label{eq:linp}\end{equation}
 The $P_{0}$ parameter has a physical basis: it represents a change
in the shot noise. If galaxies are not a Poisson sample of the matter
distribution, our power spectrum calculation will subtract the shot
noise imperfectly. Non-Poisson shot noise is a generic prediction
of the halo model \citep{2000MNRAS.318..203S}: in this context, $P_{0}$
is an approximation to the one-halo contribution to the power spectrum.
This model was proposed by \citet{2001MNRAS.325.1359S,2006APh....25..172S,2007MNRAS.375.1329G}
and has been explored by, e.g., \citet{2008JCAP...07..017H,2009MNRAS.392..682C}.
We refer to it as the P-model. \label{P} 
\item We can do the same for model~\ref{lin}, giving \begin{equation}
P_{\mathrm{gal}}\left(k\right)=b^{2}P_{\mathrm{hf}}\left(k\right)+P_{0}.\label{eq:smithp}\end{equation}
 \label{Psmith} 
\item As an alternative to the P-model, we also test the Q-model put forth
by \citet{2005MNRAS.362..505C.good}, which adds a different free
parameter $Q$ to {}``correct'' for unknown scale-dependent galaxy
bias and redshift space distortions: \begin{equation}
P_{\mathrm{gal}}\left(k\right)=b^{2}\frac{1+Qk^{2}}{1+1.4k}P_{\mathrm{lin}}\left(k\right).\label{eq:q}\end{equation}
\citet{2005MNRAS.362..505C.good} test and calibrate this model with
N-body simulations populated by a semi-analytic galaxy formation model
under the assumption of massless neutrinos -- its validity in the
massive neutrino case has not been tested in detail. We include it
here to facilitate comparison with other analyses that adopt this
model, e.g., \citet{2006PhRvD..74l3507T.good}.\label{Q} 
\item It is not clear that either the P-model or the Q-model correctly accounts
for the damping of the Baryon Acoustic Oscillations (BAO) by the non-linear
structure growth and redshift-space distortions. In order to confirm
that this is not a big effect for neutrino mass constraints, we also
consider a model with \begin{equation}
P_{\mathrm{gal}}\left(k\right)=b^{2}\frac{1+Qk^{2}}{1+1.4k}P_{\mathrm{dw}}\left(k\right),\label{eq:qdw}\end{equation}
 where $P_{\mathrm{dw}}\left(k\right)$ is the {}``dewiggled'' power
spectrum calculated in the manner of \citet{2006PhRvD..74l3507T.good}
that accounts for the non-linear damping of the BAO: \begin{equation}
P_{\mathrm{dw}}\left(k\right)=W\left(k\right)P_{\mathrm{lin}}\left(k\right)+\left[1-W\left(k\right)\right]P_{\mathrm{nowiggle}}\left(k\right),\label{eq:dewiggled}\end{equation}
 where $W\left(k\right)\equiv\mathbf{e^{-\left(k/k_{*}\right)^{2}/2}}$
and $P_{\mathrm{nowiggle}}\left(k\right)$ represents a smooth power
spectrum without BAO that we calculate using a smooth b-spline fit
to $k^{1.5}P_{\mathrm{lin}}\left(k\right)$ with eight nodes equally
spaced in $\log\left(k\right)$ from $k=0.007$ to $k=0.7\,\mathrm{Mpc^{-1}}$.
This is similar to the method used in \citet{2010MNRAS.404...60R.good}
for massive neutrino models. Following \citet{2006PhRvD..74l3507T.good}
we define $k_{*}\equiv\sigma_{\perp}^{-2/3}\sigma_{\parallel}^{-1/3}\left(A_{s}/0.6841\right)^{-1/2}$
with $\sigma_{\perp}$ and $\sigma_{\parallel}$ given by equations~(12)~and~(13)
in \citet{2007ApJ...664..660E}.\label{Qdw} 
\item We now consider a model where we replace the linear matter power spectrum
in \ref{lin}, with $P_{\mathrm{pt}}\left(k\right)$ given by equation~\eqref{eq:saito}:
\begin{equation}
P_{\mathrm{gal}}\left(k\right)=b^{2}P_{\mathrm{pt}}\left(k\right),\label{eq:linsaito}\end{equation}
 \label{saito} 
\item We can do the same for the P-model \ref{P} \begin{equation}
P_{\mathrm{gal}}\left(k\right)=b^{2}P_{\mathrm{pt}}\left(k\right)+P_{0}.\label{eq:saitop}\end{equation}
 \label{saitop} 
\item For our next model, we use an extension of $P_{\mathrm{pt}}\left(k\right)$
to include nonlinear galaxy bias. Starting from the assumption of
local deterministic nonlinear bias, \citet{2009PhRvD..80h3528S} derive
an expression for $P_{\mathrm{gal}}\left(k\right)$ using the perturbation
theory methods from \citet{2006PhRvD..74j3512M,2009ApJ...691..569J}
involving one new free parameter $b_{2}$: \begin{equation}
P_{\mathrm{gal}}\left(k\right)=b^{2}\left[P_{\mathrm{pt}}\left(k\right)+b_{2}P_{b2,\delta}\left(k\right)+b_{2}^{2}P_{b22}\left(k\right)\right]+P_{0},\label{eq:pt}\end{equation}
 where $P_{b2,\delta}\left(k\right)$ and $P_{b22}\left(k\right)$
are given by equation~(41) in \citet{2009PhRvD..80h3528S} and $b_{2}$
is a third free parameter allowed to take on different values for
different galaxy subsamples. We label this model as {}``PT with $b$,
$P_{0}$, $b_{2}$'' in our results.\label{pt} 
\item Next we consider extending the perturbations theory analysis into
redshift-space following \citet{2009PhRvD..80l3503T} and applying
the redshift space distortion model of \citet{2004PhRvD..70h3007S}.
According to \citet{2004PhRvD..70h3007S}, the redshift space power
spectrum of matter can be modeled by \begin{eqnarray}
\lefteqn{P_{\mathrm{zdist}}\left(k,\mu\right)=\left[P_{\delta\delta}\left(k\right)+2f\mu^{2}P_{\delta\theta}\left(k\right)+f^{2}\mu^{4}P_{\theta\theta}\left(k\right)\right]}\nonumber \\
 &  & \times\exp\left(-\left(f\mu k\sigma_{v}\right)^{2}\right),\label{eq:zdist}\end{eqnarray}
 where $f\equiv d\ln D/d\ln a$ ($a$ being the scale factor and $D$
being the growth factor) is the linear growth rate , $\mu\equiv\hat{\bm{k}}\cdot\hat{\bm{z}}$
is the cosine of the angle between the wavevector $\bm{k}$ and the
line of sight $\bm{z}$. $P_{\delta\delta}\left(k\right)$ and $P_{\theta\theta}\left(k\right)$
are nonlinear density and velocity power spectra, and $P_{\delta\theta}\left(k\right)$
is the density-velocity cross-power spectrum, all calculated with
one-loop perturbation theory as defined in equations~(63-65) in \citet{2004PhRvD..70h3007S}.
The one-dimensional velocity dispersion $\sigma_{v}$ is defined as
\begin{equation}
\sigma_{v}^{2}\equiv\frac{1}{3}\int\frac{d^{3}\bm{q}}{\left(2\pi\right)^{3}}\frac{P_{\theta\theta}\left(q\right)}{q^{2}}.\label{eq:sigmav}\end{equation}
 Following \citet{2009PhRvD..80h3528S}, we calculate $P_{\delta\delta}\left(k\right)$,
$P_{\theta\theta}\left(k\right)$ and $P_{\delta\theta}\left(k\right)$
in the same manner as in equation~\eqref{eq:saito}: we apply the
one-loop corrections to the cold dark matter plus baryon portion of
the power spectrum and leave the neutrino power spectrum linear. Thus
$P_{\delta\delta,\mathrm{pt}}\left(k\right)=P_{\mathrm{pt}}\left(k\right)$
as defined in equation~\eqref{eq:saito}, and $P_{\theta\theta,\mathrm{pt}}\left(k\right)$
and $P_{\delta\theta,\mathrm{pt}}\left(k\right)$ are defined analogously.

To model the angle-averaged redshift space power spectrum, which is
what our data actually represents, we integrate equation~\eqref{eq:zdist}
over $\mu$: \begin{eqnarray}
\lefteqn{P_{\mathrm{zdist}}\left(k\right)=A_{0}\left(k\right)P_{\mathrm{pt}}\left(k\right)+2fA_{2}\left(k\right)P_{\delta\theta,\mathrm{pt}}\left(k\right)}\nonumber \\
 &  & +f^{2}A_{4}\left(k\right)P_{\theta\theta,\mathrm{pt}}\left(k\right)\label{eq:zdist_int}\end{eqnarray}
 where $A_{n}\left(k\right)$ are moments of the Gaussian term in
equation~\eqref{eq:saito}: \begin{equation}
A_{n}\left(k\right)\equiv\int_{-1}^{1}d\mu\mu^{n}\exp\left(-\left(f\mu k\sigma_{v}\right)^{2}\right)\label{eq:Ak}\end{equation}
 We model the galaxy power spectrum in redshift space with linear
bias by using bias $b$ for the density terms and no bias for the
velocity terms, giving \begin{eqnarray}
\lefteqn{P_{\mathrm{gal}}\left(k\right)=A_{0}\left(k\right)b^{2}P_{\mathrm{pt}}\left(k\right)+2fA_{2}\left(k\right)bP_{\delta\theta,\mathrm{pt}}\left(k\right)}\nonumber \\
 &  & +f^{2}A_{4}\left(k\right)P_{\theta\theta,\mathrm{pt}}\left(k\right)\label{eq:linz}\end{eqnarray}
 This model is labeled as {}``PT with $b$, zdist''. \label{linz}

\item The P-model under this redshift space distortion model becomes \begin{eqnarray}
\lefteqn{P_{\mathrm{gal}}\left(k\right)=A_{0}\left(k\right)\left[b^{2}P_{\mathrm{pt}}\left(k\right)+P_{0}\right]}\nonumber \\
 &  & +2fA_{2}\left(k\right)bP_{\delta\theta,\mathrm{pt}}\left(k\right)+f^{2}A_{4}\left(k\right)P_{\theta\theta,\mathrm{pt}}\left(k\right)\label{eq:Pz}\end{eqnarray}
 which we label as {}``PT with $b$, $P_{0}$, zdist''. \label{Pz} 
\item Finally, we combine the nonlinear bias model of equation~\eqref{eq:pt}
with the redshift space distortion model to give \begin{eqnarray}
\lefteqn{P_{\mathrm{gal}}\left(k\right)=A_{0}\left(k\right)\left\{ b^{2}\left[P_{\mathrm{pt}}\left(k\right)+b_{2}P_{b2,\delta}\left(k\right)\right.\right.}\nonumber \\
 &  & \left.\left.+b_{2}^{2}P_{b22}\left(k\right)\right]+P_{0}\right\} +2A_{2}\left(k\right)b\left[fP_{\delta\theta,\mathrm{pt}}\left(k\right)\right.\nonumber \\
 &  & \left.+b_{2}P_{b2v}\left(k\right)\right]+f^{2}A_{4}\left(k\right)P_{\theta\theta,\mathrm{pt}}\left(k\right),\label{eq:ptz}\end{eqnarray}
 which we label as {}``PT with $b$, $P_{0}$, $b_{2}$, zdist''.
\label{ptz} $P_{b2v}\left(k\right)$ is defined analogously to $P_{b2,\delta}\left(k\right)$
by replacing the density kernel with the velocity kernel in equation~(41)
in \citet{2009PhRvD..80h3528S}. 
\end{enumerate}
We calculate all of the various power spectra in equations~\eqref{eq:linz},~\eqref{eq:Pz},~and~\eqref{eq:ptz}
using a modified version of \noun{camb} provided by Saito (private
communication). Note that our redshift space distortion model does
not add any new free parameters: $P_{\mathrm{zdist}}\left(k\right)$
is fully defined by equation~\eqref{eq:zdist_int} once $P_{\mathrm{lin}}\left(k\right)$
has been determined.


Models \ref{lin}, \ref{smith}, \ref{saito} and \ref{linz} have
one free bias parameter $b$, models \ref{P}, \ref{Psmith}, \ref{Q},
\ref{Qdw}, \ref{saitop} and \ref{Pz} have two free bias parameters
($b$ and either $P_{0}$ or $Q$), and models \ref{pt} and \ref{ptz}
have three free bias parameters ($b$, $P_{0}$ and $b_{2}$). These
free parameters are all allowed to take on different values for the
red and blue galaxy subsamples.

To illustrate some of the main features of these models, we plot a
subset of them with our power spectrum data in Fig.~\ref{fig:pktests}\@.
For all but one of the curves plotted, we use the parameter set from
the $\Lambda$CDM+$\sum m_{\nu}$ MCMC chain that gives the maximum
likelihood for WMAP5 only: $\bm{\theta}^{i}\equiv$( $\Omega_{\Lambda}=0.7263$,
$\Omega_{b}/\Omega_{m}=0.1666$, $h=0.7050$, $\sigma_{8}=0.8065$,
$n_{s}=0.9580$, $\sum m_{\nu}=0.0287\,\mathrm{eV}$). The remaining
curve is plotted with the parameter set from the MCMC chain giving
the maximum likelihood the all-galaxy $P\left(k\right)$ data using
the $b^{2}P_{\mathrm{pt}}\left(k\right)$ model: $\bm{\theta}^{i}\equiv$($\Omega_{\Lambda}=0.5168$,
$\Omega_{b}/\Omega_{m}=0.1392$, $h=0.5596$, $\sigma_{8}=0.5459$,
$n_{s}=0.9140$, $\sum m_{\nu}=1.34\,\mathrm{eV}$).

Figure~\ref{fig:pktests} shows that model \ref{linz} (PT with $b$,
zdist) closely mimics model \ref{P} (the P-model) for the maximum
likelihood WMAP5 cosmology and in fact fits the data just as well
with one less free parameter. The other models plotted show different
shapes at small scales, with the$b^{2}P_{\mathrm{lin}}\left(k\right)$
model underestimating the small-scale power and $b^{2}P_{\mathrm{pt}}\left(k\right)$
overestimating it. The $b^{2}P_{\mathrm{pt}}\left(k\right)$ curve
for the high $\sum m_{\nu}$ cosmology gives a shape closer to the
data than the same model with WMAP5 cosmology because it compensates
for the excess of small-scale power by adding more massive neutrinos
that damp it out. This means that a model of bias or nonlinearity
that overestimates the small-scale power could potentially lead to
a {}``false positive'' measurement of neutino mass. The effect of
this can be seen quantitatively in our results.

\subsection{Calculating the galaxy power spectrum likelihood}

To compare our theoretical models to the power spectrum data, we convolve
our model $P_{\mathrm{gal}}\left(k\right)$ with the window function
$W\left(k_{j},\, k_{n}\right)$ discussed in Section~\ref{sub:powerspectraandmocks},
where $k_{j}$ are the central wavenumbers of the observed bandpowers
and $k_{n}$ are the wavenumbers at which we evaluate the theory power
spectra. Following \citet{2010MNRAS.404...60R.good}, we define the
convolved theory power spectrum by

\begin{eqnarray}
\lefteqn{P_{\mathrm{gal,win}}\left(k_{j},\,\bm{\theta},\,\bm{\theta}_{b}\right)=}\nonumber \\
 &  & \sum_{n}W\left(k_{j},\, k_{n}\right)P_{\mathrm{gal}}\left(k_{n}/a_{\mathrm{scl}},\,\bm{\theta},\,\bm{\theta}_{b}\right)/a_{\mathrm{scl}}^{3}-W\left(k_{j},\,0\right),\label{eq:winpk}\end{eqnarray}
 where $\bm{\theta}$ are the cosmological parameters defined in Section~\ref{sub:cmb}
and $\bm{\theta}_{b}$ are the galaxy bias parameters for the model
being used: $b$ and possibly $b_{2}$ and $P_{0}$ or $Q$.

Equation~\eqref{eq:winpk} includes two very small adjustments: the
$W\left(k_{j},\,0\right)$ term accounts for the integral constraint
issue from estimating the average galaxy density from the sample itself
\citep{2007ApJ...657..645P,2010MNRAS.404...60R.good}, and the $a_{\mathrm{scl}}$
factors adjust for the fact that we use one fiducial cosmology to
convert our galaxy redshifts into distances rather than recalculating
the distances for each cosmology \citep{2006PhRvD..74l3507T.good,2010MNRAS.404...60R.good},
with\begin{equation}
a_{\mathrm{scl}}=\left[\frac{D_{A}^{2}\left(z_{\mathrm{med}}\right)H_{\mathrm{fid}}\left(z_{\mathrm{med}}\right)}{[D_{A}^{\mathrm{fid}}\left(z_{\mathrm{med}}\right)]^{2}H\left(z_{\mathrm{med}}\right)}\right]^{1/3},\label{eq:ascl}\end{equation}
 where $D_{A}$ and $H$ are the angular diameter distance and Hubble
parameter for the cosmology defined by $\mathrm{\bm{\theta}}$, $D_{A}^{\mathrm{fid}}$
and $H_{\mathrm{fid}}$ are for the fiducial cosmology, and $z_{\mathrm{med}}$
is the median redshift of the galaxy sample.

The likelihood was calculated assuming that the power spectrum data
are distributed as a multi-variate Gaussian distribution:\begin{equation}
-2\ln L\left(\bm{\theta},\,\bm{\theta}_{b}\right)=\chi^{2}\left(\bm{\theta},\,\bm{\theta}_{b}\right)=\sum_{jk}\Delta_{j}C_{jk}^{-1}\Delta_{k},\label{eq:chisqr}\end{equation}
 where $\Delta_{j}\equiv\left[\hat{P}_{\mathrm{gal}}\left(k_{j}\right)-P_{\mathrm{gal,win}}\left(k_{j},\,\bm{\theta},\,\bm{\theta}_{b}\right)\right]$,
$\hat{P}_{\mathrm{gal}}\left(k_{j}\right)$ is the measured galaxy
power spectrum, and $C_{jk}$ is the covariance matrix calculated
as described in Section~\ref{sub:powerspectraandmocks}.

Our final galaxy power spectrum likelihood is defined by marginalizing
over the bias parameters $\bm{\theta}_{b}$ by integrating the likelihood
over the prior distribution $\pi\left(\bm{\theta}_{b}\right)$: \textbf{\begin{equation}
L_{P\left(k\right)}\left(\bm{\theta}\right)\equiv\int d^{n}\bm{\theta}_{b}L\left(\bm{\theta},\,\bm{\theta}_{b}\right)\pi\left(\bm{\theta}_{b}\right).\label{eq:marglike}\end{equation}
}  We use a flat prior on $b^{2}$ from 0 to $+\infty$ and flat priors
on $P_{0}$, $b_{2}$, and $Q$ ranging from $-\infty$ to $+\infty$.
We perform this marginalization analytically where possible, which
is the case for most of our models. For models \ref{pt} and \ref{ptz},
we have to integrate over $b_{2}$ numerically, and for models \ref{linz},
\ref{Pz}, and \ref{ptz} (the ones including redshift space distortion)
we must do the integral over $b^{2}$ numerically as well.

\subsection{Importance sampling}

\label{sub:Importance-sampling}

Because we have to evaluate the neutrino mass limit for several different
power spectrum models and $k$ cutoff values, we reduce the computational
requirement by using the technique of importance sampling \citep{2002PhRvD..66j3511L}:
we use a fixed WMAP MCMC chain and reweight the chain entries according
to the revised likelihood values calculated for the different galaxy
power spectrum fits. Given an MCMC chain of parameter values $\bm{\theta}$
drawn from a likelihood distribution $L$, one can compute parameter
constraints relative to a similar distribution $L^{\prime}$ by re-weighting
the sample according to the likelihood ratios: \begin{equation}
w_{L^{\prime}}^{i}=\frac{L^{\prime}\left(\bm{\theta}^{i}\right)}{L\left(\bm{\theta}^{i}\right)}w_{L}^{i}.\label{eq:imp_samp}\end{equation}
 Since the CMB measurement and the galaxy power spectrum measurement
are independent experiments, we can find the combined likelihood simply
by multiplying the two likelihoods together: \begin{equation}
L_{\mathrm{CMB}+P(k)}^{i}=L_{P(k)}\left(\bm{\theta}^{i}\right)L_{\mathrm{CMB}}^{i}\left(\bm{\theta}^{i}\right).\label{eq:imp_like}\end{equation}
 Thus the WMAP MCMC chain can be reweighted by \begin{equation}
w_{\mathrm{CMB}+P(k)}^{i}=L_{P(k)}^{i}\left(\bm{\theta}^{i}\right)w_{\mathrm{CMB}}^{i},\label{eq:imp_weight}\end{equation}
 so the 95\% confidence upper limit on $\sum m_{\nu}$ is the value
of $M_{\nu}^{\mathrm{lim}}$ that satisfies \begin{equation}
.95=\frac{\sum_{M_{\nu}^{i}<M_{\nu}^{\mathrm{lim}}}L_{P(k)}^{i}\left(\bm{\theta}^{i}\right)w_{\mathrm{CMB}}^{i}}{\sum_{i=1}^{N}L_{P(k)}^{i}\left(\bm{\theta}^{i}\right)w_{\mathrm{CMB}}^{i}}.\label{eq:mulim}\end{equation}
 As a test of the importance sampling method, we applied it to the
dark matter halo power spectrum $\hat{P}_{halo}(k)$ $ $derived from
the DR7 Luminous Red Galaxies using the likelihood code of \citet{2010MNRAS.404...60R.good}
to evaluate $L_{P(k)}$ Our importance sampling method yields $\sum m_{\nu}<0.64\,\mathrm{eV}$
(95\% CL) for WMAP5+$\hat{P}_{halo}(k)$, which agrees within 3\%
to the result from \citet{2010MNRAS.404...60R.good} of $\sum m_{\nu}<0.62\,\mathrm{eV}$.
We therefore conclude that importance sampling is sufficiently robust
for the purposes of our analysis in this paper.

To summarize, the steps to calculate the neutrino mass limit from
WMAP5 + galaxy power spectrum are:

\begin{enumerate}
\item For each entry of cosmological parameter values in the MCMC chain,
calculate a linear power spectrum $P_{\mathrm{lin}}\left(k\right)$
with those parameters using \noun{camb}. 
\item Choose a model for nonlinear evolution and bias, as described in $\S$\ref{sub:bias}. 
\item Calculate likelihood of this model power spectrum using a measured
galaxy power spectrum. 
\item Marginalize over the bias parameters by integrating the resulting
likelihood analytically or numerically. 
\item Use importance sampling \citep{2002PhRvD..66j3511L} to reweight the
entry in the WMAP5 MCMC chain by the likelihood from the galaxy power
spectrum:\begin{equation}
w_{\mathrm{CMB}+P(k)}^{i}=L_{P(k)}^{i}w_{\mathrm{CMB}}^{i}\label{eq:imp_weight2}\end{equation}

\item Repeat this for each entry in WMAP5 MCMC chain, and then calculate
the neutrino mass limit using histograms of the MCMC chain weighted
by $w_{\mathrm{CMB}+P(k)}$. 
\end{enumerate}

\section{Results}

\label{sec:results}

\subsection{Neutrino mass constraints}

\begin{figure*}
\includegraphics[height=0.9\textheight]{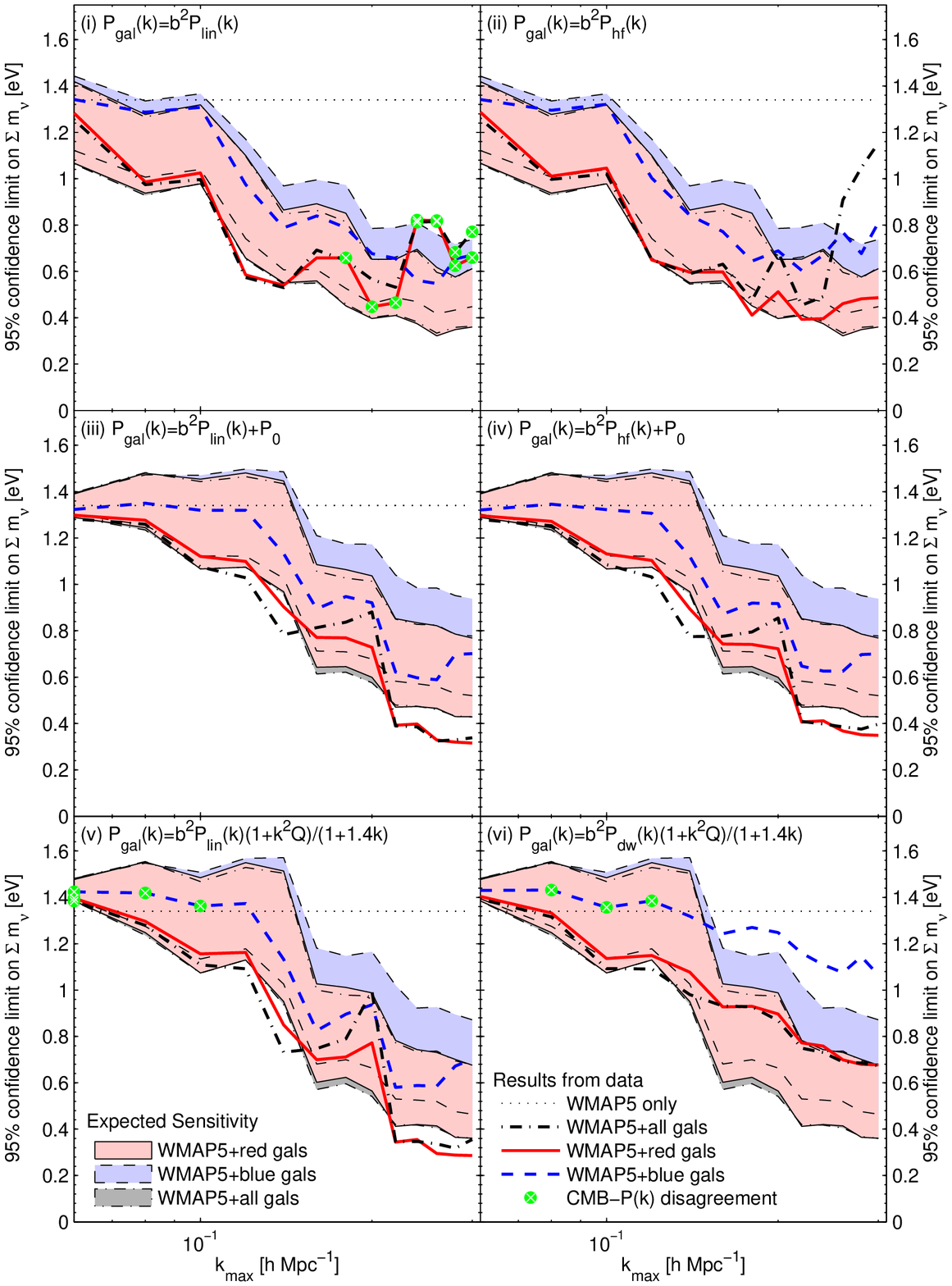}

\caption{\label{fig:nulims1}95\% CL limits on $\sum m_{\nu}$ as determined
by WMAP5 plus the SDSS-II DR7 galaxy power spectrum up to a maximum
wavenumber $k_{\mathrm{max}}$ for bias models frequently used in
the literature, e.g. \protect\citet{2002PhRvL..89f1301E.good,2005MNRAS.362..505C.good,2004PhRvD..69j3501T.good,2006PhRvD..74l3507T.good,2008JCAP...07..017H}.
Expected sensitivity ranges show the mean limit $M_{\nu}^{\mathrm{lim}}$
$\pm$ one standard deviation measured from 200 mock galaxy catalogs
and represent what we would expect to measure for the neutrino mass
limit assuming that $\sum m_{\nu}=0$ and we understood the bias perfectly.
Green points indicate cases where the CMB and galaxy datasets may
potentially be inconsistent with each other -- these points should
be regarded with caution. Using data at larger $k$ values leads to
tighter limits on $\sum m_{\nu}$ but makes the results increasingly
model-dependent. }

\end{figure*}

\begin{figure*}
\includegraphics[height=0.9\textheight]{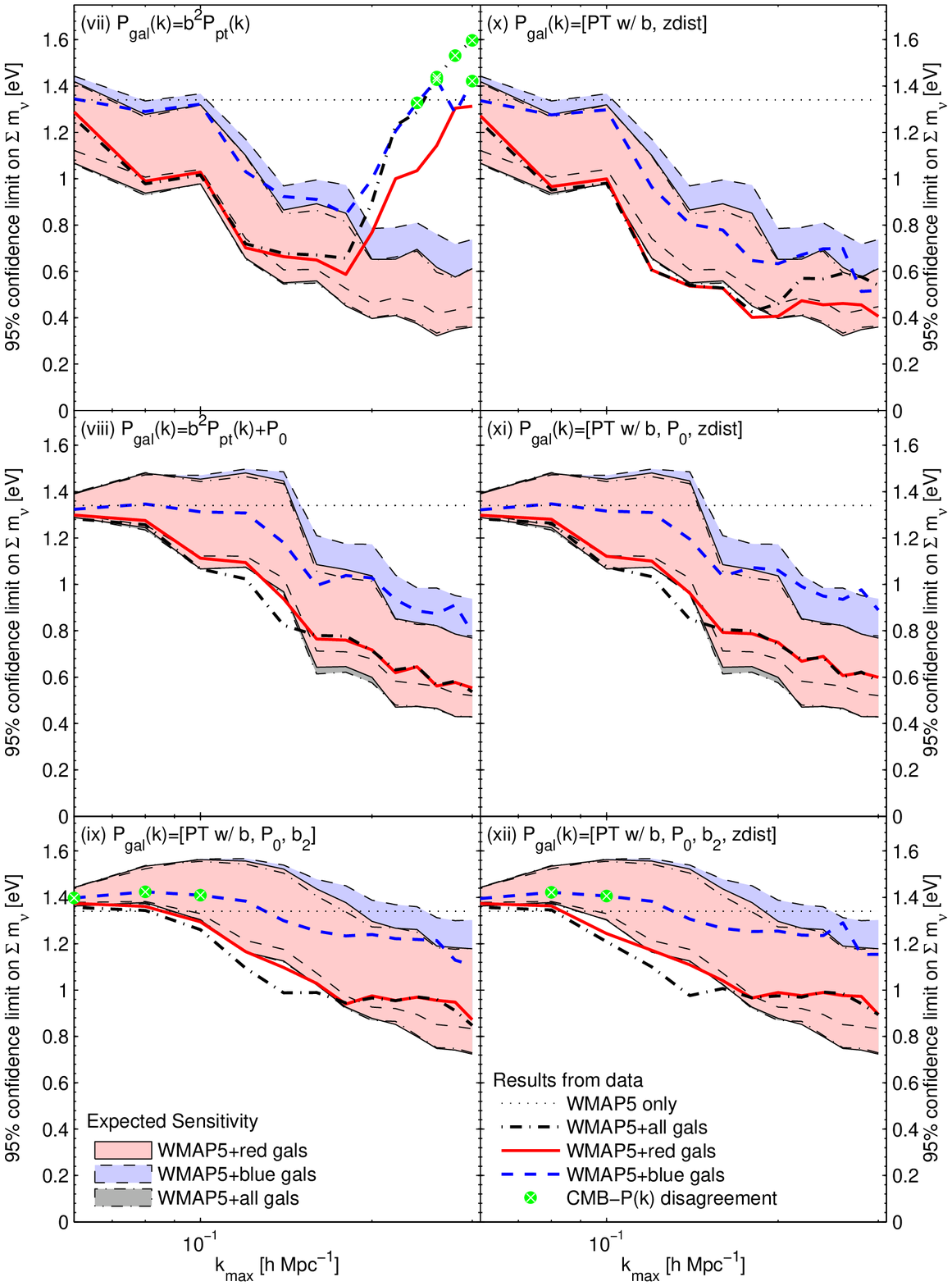}

\caption{\label{fig:nulims2}95\% CL limits on $\sum m_{\nu}$ as determined
by WMAP5 plus the SDSS-II DR7 galaxy power spectrum up to a maximum
wavenumber $k_{\mathrm{max}}$ for bias models based on perturbation
theory results from \protect\citet{2004PhRvD..70h3007S,2008PhRvL.100s1301S,2009PhRvD..80l3503T}.
Expected sensitivity ranges show the mean limit $M_{\nu}^{\mathrm{lim}}$
$\pm$ one standard deviation measured from 200 mock galaxy catalogs
and represent what we would expect to measure for the neutrino mass
limit assuming that $\sum m_{\nu}=0$ and we understood the bias perfectly.
Green points indicate cases where the CMB and galaxy datasets may
potentially be inconsistent with each other -- these points should
be regarded with caution. Using data at larger $k$ values leads to
tighter limits on $\sum m_{\nu}$ but makes the results increasingly
model-dependent. }

\end{figure*}

We have performed the above procedure using the power spectra for
the full SDSS-II DR7 main sample, and for the same sample split into
red and blue galaxies. For each of the three samples, we have done
the calculation using power spectrum values for $k<k_{\mathrm{max}}$,
with $k_{\mathrm{max}}$ ranging from $0.06$ to $0.3\, h\,\mathrm{Mpc^{-1}}$
and repeated this for each of our twelve models for bias and nonlinearities.
Results for the first six models are plotted in Fig.~\ref{fig:nulims1},
and results for the six models based on \citet{2009PhRvD..80h3528S}
are plotted in Fig.~\ref{fig:nulims2}.The neutrino limits for the
all-galaxy sample typically trace the limits for the red galaxies,
indicating that the red galaxies dominate the signal - this is because
the red galaxies are more highly biased. Furthermore, the limits from
the blue galaxy sample tend to be larger than those from the red galaxies
--- this is to be expected since the measurements using the red galaxies
are intrinsically more sensitive and therefore give more stringent
limits. We also typically obtain weaker limits as we increase the
number of free parameters used to model galaxy bias and nonlinearities,
as expected.

\begin{figure*}
\includegraphics[width=1\textwidth]{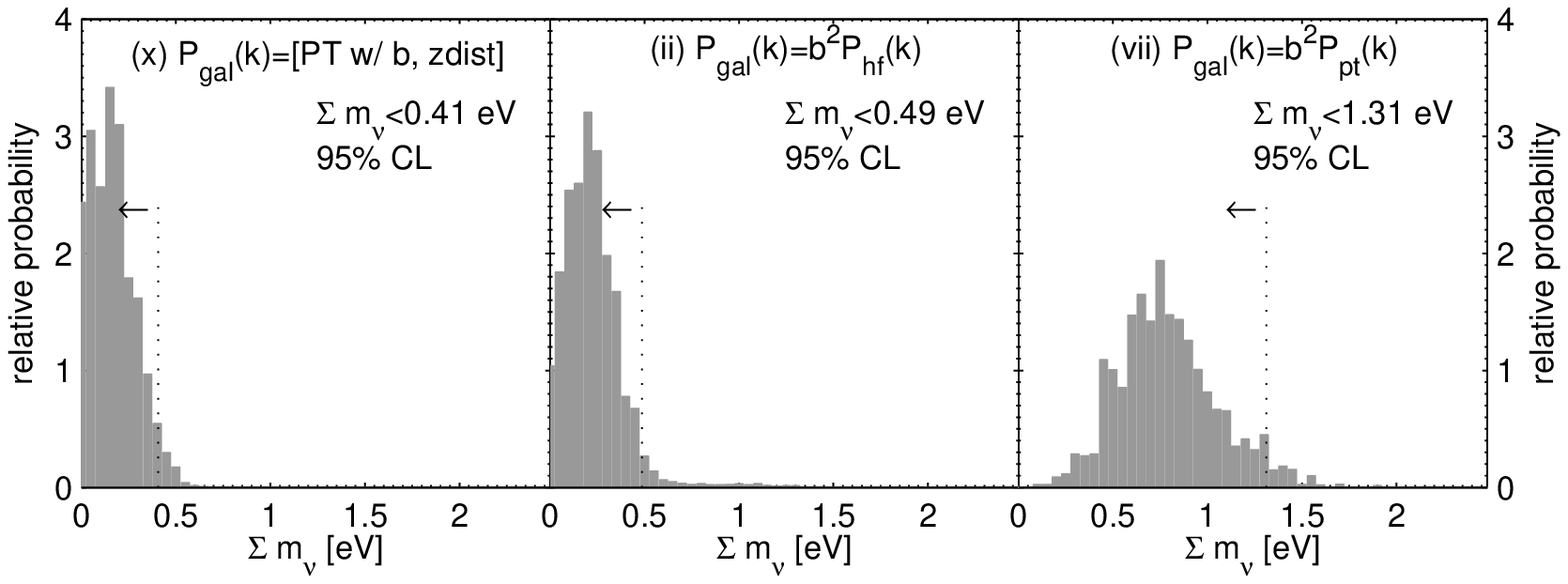}

\caption{\label{fig:mnulims_extremes}Probability distributions for $M_{\nu}$
using WMAP5 + $P(k)$ with $k<0.3\: h\,\mathrm{Mpc^{-1}}$ for the
red galaxy samples.}

\end{figure*}

One notable feature seen in Figs.~\ref{fig:nulims1}~and~\ref{fig:nulims2}
is that for the all-galaxy sample using the $b^{2}P_{\mathrm{hf}}\left(k\right)$
model and all three galaxy samples using the $b^{2}P_{\mathrm{pt}}\left(k\right)$
model, the neutrino limit increases sharply at the high end of our
$k_{\mathrm{max}}$ range. To explore what is driving this, we plot
the histograms for $k_{\mathrm{max}}=0.3\: h\,\mathrm{Mpc^{-1}}$
for the red galaxy sample using the {[}PT w/ $b$, zdist], $b^{2}P_{\mathrm{hf}}\left(k\right)$
and $b^{2}P_{\mathrm{pt}}\left(k\right)$ models in Fig.~\ref{fig:mnulims_extremes}.

The distribution for the $b^{2}P_{\mathrm{pt}}\left(k\right)$ case
clearly peaks at a non-zero neutrino mass, and a hint of a secondary
peak at non-zero neutrino mass can be seen in the $b^{2}P_{\mathrm{hf}}\left(k\right)$
histogram as well. From Fig.~\ref{fig:pktests} we can see that the
driving force behind this effect is that for models with no flexibility
in the power spectrum shape, the differences in the amount of upturn
at the high-$k$ end are compensated for by changing the cosmology
instead. For the $b^{2}P_{\mathrm{pt}}\left(k\right)$ model, the
shape of the power spectrum for the best-fit WMAP5 parameters overestimates
the data at high $k$, and increasing the neutrino mass can pull the
high-$k$ end down into better agreement with the data.

We have compared the neutrino mass limits calculated from the SDSS-II
data against those from 200 mock catalogues: the shaded regions in
Figs.~\ref{fig:nulims1} \&~\ref{fig:nulims2} contain 68\% of the
neutrino mass limits from mock catalogues. As discussed in Section~\ref{sub:powerspectraandmocks},
we would expect to measure the neutrino mass limit to be within this
expected sensitivity range if the neutrino mass were equal to zero
and we understood bias perfectly. They will therefore underestimate
the true range of limits. Because the mock catalogues are constructed
to have a power spectrum matched to $P_{\mathrm{lin}}\left(k\right)$
with linear bias and no redshift space distortions, we use the $b^{2}P_{\mathrm{lin}}\left(k\right)$
model on the mock catalogues to define the expected sensitivity range
for models \ref{lin}, \ref{smith}, \ref{saito}, and \ref{linz}
(all of the one-parameter bias models). Likewise, we use the $b^{2}P_{\mathrm{lin}}\left(k\right)+P_{0}$
model for the expected sensitivity range for models \ref{P},\ref{Psmith},
\ref{saitop}, and \ref{Pz} (all of the P-models ), the $b^{2}\left(1+Qk^{2}\right)/\left(1+1.4k\right)P_{\mathrm{lin}}\left(k\right)$
model for both of the Q-models \ref{Q} and \ref{Qdw}, and the {[}PT
w/ $b$, $P_{0}$, $b_{2}$] model for both of the $b_{2}$ models
\ref{pt} and \ref{ptz}. This correctly captures the weakening of
sensitivity due to adding more free parameters but still allows for
a good fit to the power spectrum built into the mocks. (Note that
this is not strictly true for the Q-models as $Q=0$ does not reduce
to $P_{\mathrm{lin}}\left(k\right)$; however, we expect the effects
of this to be negligible.)

Comparing the data curves to the shaded regions in Figs.~\ref{fig:nulims1}~and~\ref{fig:nulims2},
we see that the results from red and blue galaxies are generally consistent
with the limits predicted from the mock catalogues at the 1$\sigma$
level, which suggests that current neutrino mass limits from large
scale structure are sufficiently robust to differences in bias modeling.
However, the constraints from the red galaxies are systematically
better than expected, while the results from the blue galaxies are
systematically worse. We consider this difference further in Section~\ref{sec:redbluecompare}.

\begin{table*}
\begin{minipage}[c][1\totalheight]{174mm}%
\caption{\label{tab:nulim_summary}Summary of neutrino limits, calculated to
a conservative wavenumber limit $k_{\mathrm{max}}<0.12\,\hompc$,
and to a more aggressive limit $k_{\mathrm{max}}<0.2\,\hompc$. Models
are classified by the number of free parameters that are used to model
galaxy bias.}

\begin{tabular}{ll|c|llll|llll}
\hline 
\multicolumn{2}{l|||||}{Limits on $\sum m_{\nu}$ in eV at 95\% CL } & \# of free  & \multicolumn{4}{l|}{Conservative ($k_{\mathrm{max}}<0.12\,\hompc$)} & \multicolumn{4}{l}{Aggressive ($k_{\mathrm{max}}<0.2\,\hompc$)}\tabularnewline
\multicolumn{2}{l||||||}{Model } & params  & $\qquad$  & blue  & red  & all  & $\qquad$  & blue  & red  & all\tabularnewline
\hline 
\ref{lin}  & $b^{2}P_{\mathrm{lin}}\left(k\right)$  & 1  &  & 0.98  & 0.58  & 0.57  &  & 0.68  & 0.45  & 0.56\tabularnewline
\ref{smith}  & $b^{2}P_{\mathrm{hf}}\left(k\right)$  & 1  &  & 1.00  & 0.65  & 0.65  &  & 0.69  & 0.51  & 0.68\tabularnewline
\ref{P}  & $b^{2}P_{\mathrm{lin}}\left(k\right)+P_{0}$  & 2  &  & 1.32  & 1.10  & 1.03  &  & 0.92  & 0.73  & 0.88\tabularnewline
\ref{Psmith}  & $b^{2}P_{\mathrm{hf}}\left(k\right)+P_{0}$  & 2  &  & 1.31  & 1.10  & 1.03  &  & 0.92  & 0.72  & 0.85\tabularnewline
\ref{Q}  & $b^{2}\left(1+Qk^{2}\right)/\left(1+1.4k\right)P_{\mathrm{lin}}\left(k\right)$  & 2  &  & 1.37  & 1.16  & 1.09  &  & 0.94  & 0.77  & 0.99\tabularnewline
\ref{Qdw}  & $b^{2}\left(1+Qk^{2}\right)/\left(1+1.4k\right)P_{\mathrm{dw}}\left(k\right)$  & 2  &  & 1.39  & 1.15  & 1.09  &  & 1.25  & 0.90  & 0.87\tabularnewline
\ref{saito}  & $b^{2}P_{\mathrm{pt}}\left(k\right)$  & 1  &  & 1.03  & 0.70  & 0.72  &  & 1.00  & 0.77  & 0.90\tabularnewline
\ref{saitop}  & $b^{2}P_{\mathrm{pt}}\left(k\right)+P_{0}$  & 2  &  & 1.31  & 1.09  & 1.02  &  & 1.03  & 0.72  & 0.71\tabularnewline
\ref{pt}  & PT with $b$, $P_{0}$, $b_{2}$  & 3  &  & 1.38  & 1.17  & 1.09  &  & 1.24  & 0.97  & 0.97\tabularnewline
\ref{linz}  & PT with $b$, zdist  & 1  &  & 0.96  & 0.61  & 0.61  &  & 0.63  & 0.41  & 0.46\tabularnewline
\ref{Pz}  & PT with $b$, $P_{0}$, zdist  & 2  &  & 1.31  & 1.10  & 1.03  &  & 1.06  & 0.75  & 0.74\tabularnewline
\ref{ptz}  & PT with $b$, $P_{0}$, $b_{2}$, zdist  & 3  &  & 1.38  & 1.17  & 1.10  &  & 1.25  & 0.99  & 0.97\tabularnewline
\hline
\end{tabular}%
\end{minipage}%

\end{table*}

We show our 95\% confidence limits on the summed neutrino mass for
each model Table~\ref{tab:nulim_summary}, using both conservative
($k_{\mathrm{max}}<0.12\,\hompc$) and aggressive ranges of scale
($k_{\mathrm{max}}<0.2\,\hompc$). The effect of the number of free
parameters in the model (excluding the effect of neutrinos) clearly
has a strong effect on the recovered constraint. This is discussed
further in Section~\ref{sec:conclusions}.

\subsection{Consistency between CMB and LSS data}

When combining results from independent datasets such as our combination
of CMB and galaxy power spectrum data, it is important to check that
they are consistent with each other under the model. If the galaxy
power spectrum is not fit well by any set of parameters that has a
reasonably high likelihood with respect to the CMB data, results obtained
by combining them will be meaningless. \citet{2006PhRvD..73f7302M}
details a method to determine whether multiple datasets are consistent:
compute the ratio of Bayesian evidences for two hypotheses $H_{0}$
and $H_{1}$, where $H_{0}$ is {}``There is one set of parameters
defining one global model that describes all datasets,'' and $H_{1}$
is {}``There is a different set of parameters that describes each
dataset.''

For two independent datasets $\bm{d}_{1}$ and $\bm{d}_{2}$, this
ratio is (see equation~(3) of \citet{2006PhRvD..73f7302M}) \begin{equation}
R\equiv\frac{\mathrm{Pr}\left(\bm{d}_{1},\,\bm{d}_{2}|H_{0}\right)}{\mathrm{Pr}\left(\bm{d}_{1}|H_{1}\right)\mathrm{Pr}\left(\bm{d}_{2}|H_{1}\right)}.\label{eq:R_marshall}\end{equation}
 Because calculating the evidence requires integrating over the prior
distribution for all of the parameters of the model, it is computationally
difficult to calculate. Motivated by \citet{1995KassRaftery}, we
define an approximation to the evidence that can be easily computed
from an MCMC chain:\begin{equation}
\hat{\mathrm{Pr}}\left(\bm{d}_{1}\right)\equiv\left\{ \frac{\sum_{i=1}^{N}w_{1}^{i}\left[\mathrm{Pr}\left(\bm{d}_{1}|\bm{\theta}^{i}\right)\right]^{-1}}{\sum_{i=1}^{N}w_{1}^{i}}\right\} ^{-1}\label{eq:evidence_KR}\end{equation}
 This is the harmonic mean of the likelihood values in the MCMC chain
and can be derived following \citet{1995KassRaftery} by using importance
sampling: since the MCMC chain samples the posterior distribution,
we can approximate an integral of $\mathrm{Pr}\left(\bm{d}_{1}|\bm{\theta}^{i}\right)$
over the prior distribution with a sum over the MCMC chain by reweighting
using equation~\eqref{eq:imp_samp} with $L^{\prime}$ equal to the
prior distribution $\pi\left(\bm{\theta}^{i}\right)$ and $L$ equal
to the posterior distribution $\mathrm{Pr}\left(\bm{\theta}^{i}|\bm{d}_{1}\right)=\mathrm{Pr}\left(\bm{d}_{1}|\bm{\theta}^{i}\right)\pi\left(\bm{\theta}^{i}\right)/\mathrm{Pr}\left(\bm{d}_{1}\right)$.
\citep{1995KassRaftery} indicate that while this quantity can be
unstable in some cases, it is typically accurate enough for making
rough categorizations, which is how we use it here.

Following this same idea, we derive an approximation for $R$ in the
case where we have an MCMC chain for $\bm{d}_{1}$ and are using importance
sampling to combine $\bm{d}_{1}$ and $\bm{d}_{2}$, as we describe
in \S\eqref{sub:Importance-sampling} for $\bm{d}_{1}$=WMAP5 and
$\bm{d}_{2}=P_{\mathrm{gal}}(k)$ :\begin{equation}
\hat{R}\equiv\frac{\left\{ \sum_{i=1}^{N}w_{1+2}^{i}\right\} /\left\{ \sum_{i=1}^{N}w_{1+2}^{i}\left[\mathrm{Pr}\left(\bm{d}_{1}|\bm{\theta}^{i}\right)\right]^{-1}\right\} }{\left\{ \sum_{i=1}^{N}w_{1}^{i}\right\} /\left\{ \sum_{i=1}^{N}w_{1}^{i}\left[\mathrm{Pr}\left(\bm{d}_{1}|\bm{\theta}^{i}\right)\right]^{-1}\right\} },\label{eq:R_approx}\end{equation}
 where $w_{1+2}^{i}$ is given, e.g., by equation~\eqref{eq:imp_weight}.
This is quite similar to the heuristic approach proposed by \citet{2002PhRvD..66j3511L}
of comparing the mean likelihood of the samples where the posterior
distributions overlap to the mean likelihood under the original posterior:
here we have simply replaced the mean by the harmonic mean. Following
\citet{2006PhRvD..73f7302M}, we interpret a value of $\hat{R}>1$
as indicating that the null hypothesis $H_{0}$ of a global parameter
set is favored over $H_{1}$ with a separate parameter set for each
data set. Conversely, $\hat{R}<1$ indicates that $H_{1}$ is favored.
Obtaining a value of $\hat{R}<1$ does not automatically guarantee
that the datasets disagree, but it is an indicator that a possible
mismatch between the datasets should be investigated further. 

In this analysis we use $\hat{R}$ as a conservative {}``warning
label'': we mark models where $\hat{R}<1$ as an indication that
we should proceed with caution in interpreting the results since there
may be disagreement between WMAP5 and $P_{\mathrm{gal}}(k)$ under
these models. These points are clearly marked in Figs.~\ref{fig:nulims1}
\&~\ref{fig:nulims2}. In particular, this warning label applies
to the points at the high $k_{\mathrm{max}}$ end of the $b^{2}P_{\mathrm{pt}}\left(k\right)$
results where the measured limits from the combined datasets are higher
than the WMAP5 only limit. This suggests that applying such a consistency
check to future datasets could avoid a spurious detection of neutrino
mass caused by an incorrect bias model.

\subsection{Comparison between results from blue and red galaxies}

\label{sec:redbluecompare}

\begin{figure*}
\includegraphics[width=1\textwidth]{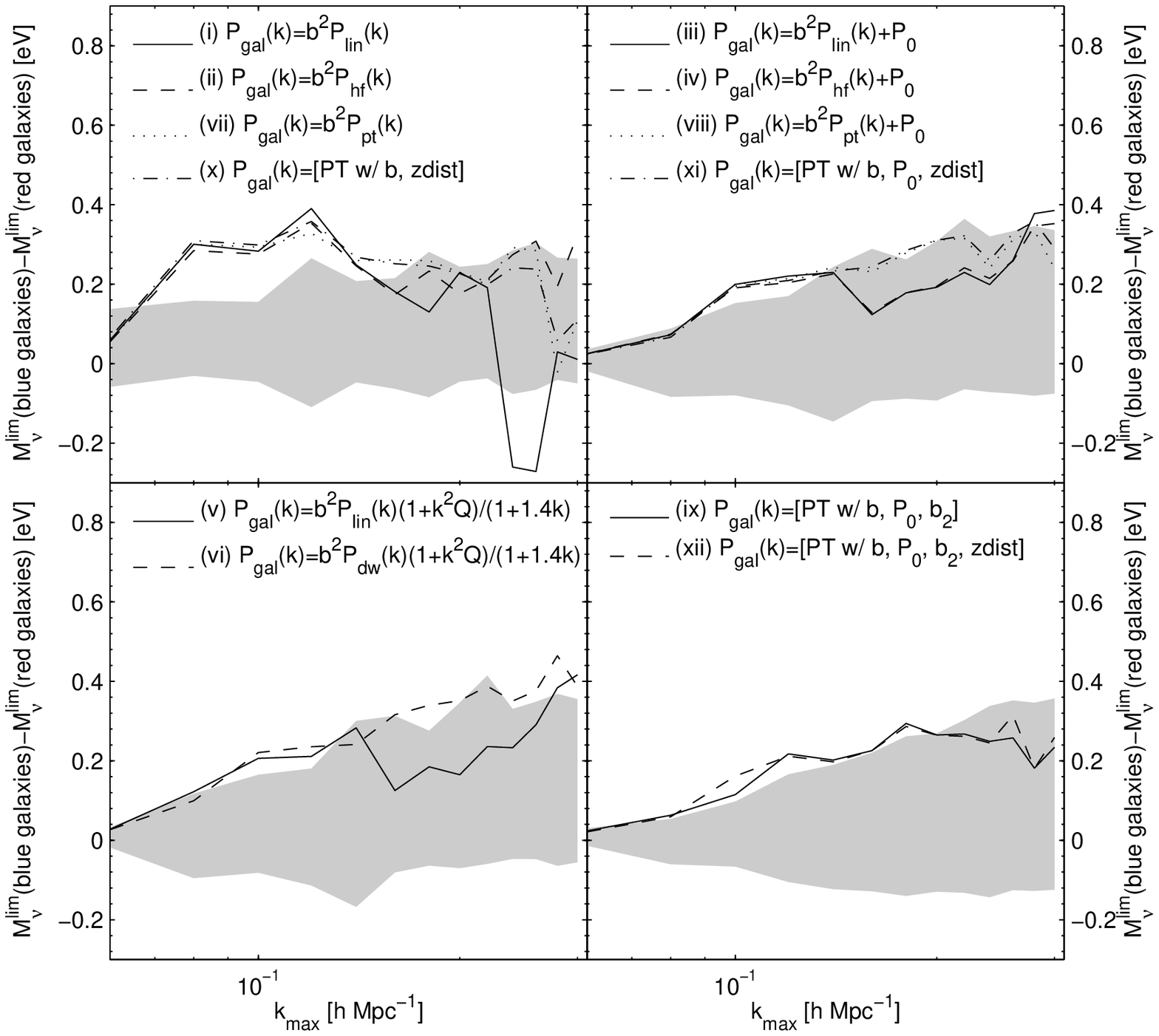}

\caption{\label{fig:blueminusred}Difference between $M_{\nu}^{\mathrm{lim}}$
as determined by WMAP5 plus the SDSS-II DR7 red and blue galaxy power
spectra. Shaded region shows the mean $M_{\nu}^{\mathrm{lim}}(\mathrm{blue})-M_{\nu}^{\mathrm{lim}}(\mathrm{red})$
$\pm$ one standard deviation measured from 200 mock galaxy catalogs
and represent what we would expect to measure assuming that $M_{\nu}=0$
and we understood the bias perfectly. }

\end{figure*}

To display these effects more clearly, and include the correlations
between the blue and red galaxy samples (caused by the overlapping
volume covered), we plot the difference between the neutrino mass
limits from red and blue galaxies as compared to the expected range
from the mock catalogs in Fig.~\ref{fig:blueminusred}. To define
the expected range, we compute the difference between the neutrino
limits derived from the mock blue and red galaxy samples drawn from
the same mock LN density distribution. Then we find the mean and standard
deviation of this difference over the 200 mocks, and plot the expected
range as the mean $\pm$ one standard deviation. This accounts for
correlations between the red and blue galaxy samples due to being
drawn from the same underlying distribution.

Figure~\ref{fig:blueminusred} shows that the difference between
the blue and red galaxy limits is consistent at approximately the
$1\sigma$ level, compared with the expected range for nearly all
of the different bias models the entire $k_{\mathrm{max}}$ range
tested. Red galaxy limits are systematically higher than those from
the blue galaxies, even after accounting for the expected difference
in sensitivity, but the significance of this is low.

\section{Conclusions}

\label{sec:conclusions}

We have performed a quantitative analysis considering how large-scale
structure neutrino mass limits derived from galaxy surveys depend
on our ability to model non-linear effects such as galaxy bias and
redshift-space distortions. Galaxies form a bimodal distribution in
colour and can be split into red and blue galaxies. These populations
cluster in very different ways, with red galaxies being more strongly
clustered on large scales than blue \citep{2002MNRAS.332..827N.good,2004PhRvD..69j3501T.good,2005MNRAS.356..247W.good,2005MNRAS.361..415C,2008MNRAS.385.1635S},
and have a different redshift-space power spectrum shape \citep{2009MNRAS.392..682C}.
Splitting into red and blue galaxies and analysing the samples separately
is therefore an obvious way to empirically analyse the robustness
of neutrino mass limits from galaxy clustering measurements, and we
have performed this test using data from the SDSS-II DR7. To evaluate
the robustness of the derived neutrino mass limits, we compare the
limits to the expected sensitivity range calculated from mock Log-Normal
catalogues: given the accuracy with which we have measured the galaxy
power spectrum, the expected sensitivity range defines the neutrino
mass limit we expect to measure if the neutrino mass were equal to
zero and we understood bias perfectly.

Galaxy bias and redshift-space effects alter the measured galaxy power
spectrum from that of the matter power spectrum and many different
models have been suggested to account for this. To see how the choice
of model affects the results for the neutrino mass limit, we have
considered 12 of these models, described in Section~\ref{sub:bias},
that span the space of currently-used models and also include new
models \citep{2009PhRvD..80h3528S,2009PhRvD..80l3503T}.

Our models can be classified by the number of galaxy bias parameters
used: the simplest models use only one parameter, namely the constant
linear bias $b$ which simply renormalizes the power spectrum. More
complicated models use two or three parameters to account for more
complex scale-dependent effects. It is clear from the results presented
in Figs.~\ref{fig:nulims1} \&~\ref{fig:nulims2} that most one-parameter
models break down for $k_{\mathrm{max}}\simgt0.1-0.2$, giving results
beyond these limits that are inconsistent with the expected sensitivity
range. The exception is model~\ref{linz} (a combination of the perturbation
theory model of {[}\citealt{2009PhRvD..80h3528S}] and the redshift
space distortion model of {[}\citealt{2004PhRvD..70h3007S}]), which
fits impressively well with only one free parameter.

All models with two free parameters give remarkably similar results:
clearly the inclusion of the extra parameter allows all of these models
to mimic the broad features of galaxy bias and redshift-space distortions,
leading to consistent constraints. The only 2-parameter model that
stands out is the dewiggled $Q$-model, which is unable to fit the
shape of the blue galaxy power spectrum\textbf{.} Three-parameter
models give self-consistent but not very informative results: the
inclusion of another free parameter significantly degrades our ability
to see the signal of massive neutrinos. We also note that adding more
free parameters increases the possibility of degeneracy with cosmological
parameters, which could potentially bias the results for large $k$
values \citep{2010arXiv1006.4845S}.

Overall, we find constraints from the red galaxies that are slightly
stronger than expected: they are at the 1$\sigma$ edge of expected
constraints, while those for blue galaxies are slightly less restrictive
than expected, again approximately at the edge of the $1\sigma$ range
expected. In order to assess the significance of this potential discrepancy
we need to allow for correlations between the recovered red and blue
galaxy power spectra as they cover the same volume of the Universe.
We account for this using mock Log-Normal catalogues, drawing overlapping
red and blue mock samples from the same density fields. Figure~\ref{fig:blueminusred}
shows that the blue galaxy limit minus the red galaxy limit is slightly
larger than expected: the difference in the neutrino mass result between
SDSS-II red and blue galaxies is at the 1$\sigma$ limit. This matches
the results shown in Figs.~\ref{fig:nulims1} \&~\ref{fig:nulims2},
indicating that the effect of correlations between red and blue power
spectra is small.

We therefore see that, with current SDSS data, recovered neutrino
mass constraints are broadly consistent for red and blue galaxies.
With only these data we cannot tell if the intriguing differences
between blue and red galaxies are caused by noise, or if there is
an element lacking in all of the bias models we have tested. Our results
indicate that current cosmological neutrino mass constraints are generally
robust to astrophysical systematic effects, but that analyses of future
surveys will need to treat these effects carefully in order to set
convincingly tighter constraints. It is clear that being able split
a galaxy catalogue into red and blue galaxies provides an important
diagnostic test of potential systematic errors, when measuring neutrino
masses. When designing future surveys, it will be important to allow
for such tests for systematics.


\section*{Acknowledgments}

We thank Shun Saito for kindly providing his modified \noun{camb}
code, and Filipe Abdalla, Sarah Bridle, {\O}ystein Elgar{\o}y,
Angeliki Kiakotou, Donnacha Kirk, and Shaun Thomas for useful discussions.
We thank the makers of \noun{camb}\footnotemark[2] for making their
code public, SDSS team for making their data public via \texttt{http://www.sdss.org}
and the WMAP team for making their data and Monte Carlo Markov Chains
public via LAMBDA\footnotemark[1]. Support for LAMBDA is provided
by the NASA Office of Space Science. MS was supported by the National
Science Foundation under Award No. OISE-0754357 and Award No. AST-0901965.
WJP is grateful for support from the UK Science and Technology Facilities
Council, the Leverhulme Trust and the European Research Council. OL
acknowledges a Royal Society Wolfson Research Merit Award and an Erna
\& Jakob Michael Visiting Professorship at the Weizmann Institute
of Science. \bibliographystyle{mymn2e}
\bibliography{redblue,redblue1}

\end{document}